# Extraction a Formalism for Fluids with Non-Spherical Molecules based on the Cluster Expansion of the Energy Functional

S. M. Hekmatzadeh [1], G. H. Bordbar [1,2*†] and M. Moradi [1]

[1]*Physics Department, College of Sciences, Shiraz University, Shiraz 71454, Iran*

[2] *Department of Physics and Astronomy, University of Waterloo, 200 University Avenue West, Waterloo, Ontario N2L3G1, Canada*

## Abstract

In this work, based on the cluster expansion of the energy functional, we have extracted a formalism for calculation of the thermodynamic properties of fluids with non-spherical molecules. The salient feature of the extracted formalism is that it has no restrictions on the type of interaction. In fact our formalism can be employed for all types of realistic inter-particle interactions. Here, for the interaction of two anisotropic molecules, we have applied the Gay-Berne potential. Finally, we have applied this formalism to calculate some thermodynamic properties of fluid $H_2$ which shows some expected results.

---

* ghbordbar@shirazu.ac.ir

† Corresponding Author

## 1. INTRODUCTION

As it is well known that the liquids, gases and plasmas are considered in the category of fluids. In nature, there are different fluids with distinctive features. If the interaction between the molecules of a fluid has spherical symmetry, it is called simple fluids. In simple fluids with spherical molecules, e.g. inert gases, the inter-molecular interaction depends only on the inter-molecular distance $r$. In molecular fluids with non-spherical molecules, for example $H_2O$, $NH_3$ and $H_2$, the inter-molecular interaction depends also on the molecular orientations $\hat{u}_1$ and $\hat{u}_2$ in addition to $r$. Here $\hat{u}_i \equiv (\theta_i \phi_i)$ (polar angles) for linear molecules (e.g. $H_2$) and $\hat{u}_i \equiv (\theta_i \phi_i \chi_i)$ (Euler angles) for nonlinear molecules (e.g. $H_2O$) showing the orientations relative to a fixed space coordinate system. The dependence of molecular orientation leads to different behavior for molecular fluids compared to the simple fluids [1].

Understanding the inter-molecular interaction potential is one of the main foundations in the study of the structural and thermodynamic properties of a system. Interactions determine the physical and chemical properties of gases, liquids and crystals [2]. Due to the large number of particles in the systems, the shape of interaction potential is extracted by some approximations based on the phenomenological considerations. Gay-Berne potential model [3] is one of the useful inter-molecular interaction models which clearly combines anisotropy in attraction and repulsion parts of the interaction, and it is of special interests for calculation of the properties of non-spherical molecules. The Gay-Berne model is an anisotropic form of the Lennard-Jones potential. This model is usually used in computer simulation of liquid crystals [4, 5, 6, 7] and density functional theory [8, 9]. The intermolecular pair potential proposed by Gay and Berne is written as

$$U(\vec{r}, \hat{u}_i, \hat{u}_j) = 4\varepsilon(\hat{r}, \hat{u}_i, \hat{u}_j) \left\{ \left[ \frac{\sigma_0}{r - \sigma(\hat{r}, \hat{u}_i, \hat{u}_j) + \sigma_0} \right]^{12} - \left[ \frac{\sigma_0}{r - \sigma(\hat{r}, \hat{u}_i, \hat{u}_j) + \sigma_0} \right]^{6} \right\}, \quad (1)$$

where the unit vector $\hat{u}_i$ defines the orientation of main axis of molecule $i$, and $\hat{r} = \vec{r}/r$ is the unit vector along $\vec{r} = \vec{r}_2 - \vec{r}_1$. Here, $\vec{r}_1$ and $\vec{r}_2$ are the center of mass positions of molecules 1 and 2, respectively. The directional dependence of the range parameter $\sigma(\hat{r}, \hat{u}_i, \hat{u}_j)$ and the potential well depth $\varepsilon(\hat{r}, \hat{u}_i, \hat{u}_j)$ are defined as follows,

$$\sigma(\hat{r}, \hat{u}_i, \hat{u}_j) = \sigma_0 \left\{ 1 - \frac{1}{2}\chi \left[ \frac{(\hat{u}_i \cdot \hat{r} + \hat{u}_j \cdot \hat{r})^2}{1 + \chi(\hat{u}_i \cdot \hat{u}_j)} + \frac{(\hat{u}_i \cdot \hat{r} - \hat{u}_j \cdot \hat{r})^2}{1 - \chi(\hat{u}_i \cdot \hat{u}_j)} \right] \right\}^{-\frac{1}{2}}, \quad (2)$$

$$\varepsilon(\hat{r}, \hat{u}_i, \hat{u}_j) = \varepsilon_0 \varepsilon^{\nu}(\hat{u}_i, \hat{u}_j) \varepsilon'^{\mu}(\hat{r}, \hat{u}_i, \hat{u}_j), \quad (3)$$

where $\sigma_0$ and $\varepsilon_0$ are constants that represent the molecular breadth and the well depth for the cross-configuration, respectively. In above relations,

$$\varepsilon(\hat{u}_i, \hat{u}_j) = \left[1 - \chi^2(\hat{u}_i \cdot \hat{u}_j)^2\right]^{-\frac{1}{2}}, \tag{4}$$

and

$$\varepsilon'(\hat{r}, \hat{u}_i, \hat{u}_j) = 1 - \frac{1}{2}\chi' \left[\frac{(\hat{u}_i \cdot \hat{r} + \hat{u}_j \cdot \hat{r})^2}{1 + \chi'(\hat{u}_i \cdot \hat{u}_j)} + \frac{(\hat{u}_i \cdot \hat{r} - \hat{u}_j \cdot \hat{r})^2}{1 - \chi'(\hat{u}_i \cdot \hat{u}_j)}\right], \tag{5}$$

where $\chi$, the shape anisotropy parameter, is determined from the ratio of molecular length to breadth, $k$,

$$\chi = \frac{(k^2 - 1)}{(k^2 + 1)}, \tag{6}$$

and the parameter $\chi'$, determines the energy anisotropy,

$$\chi' = \frac{\left(k'^{\frac{1}{\mu}} - 1\right)}{\left(k'^{\frac{1}{\mu}} + 1\right)}. \tag{7}$$

$k'$ is the ratio of the potential well depths for the side-by-side and end-to-end configurations of the molecules. $\mu$ and $\nu$ are two modifiable parameters that give a broad collection of anisotropic potentials. The first model was selected by Gay and Berne by choosing $\mu = 2$ and $\nu = 1$. Therefore, the exact shape of potential is determined with four parameters $k$, $k'$, $\mu$ and $\nu$. When $k$ and $k'$ are equal to unity, the Gay-Berne potential reduces to the Lennard-Jones potential. In position $r = \sigma(\hat{r}, \hat{u}_i, \hat{u}_j)$, the repulsion and attraction parts cancel each other, and the potential is zero.

In the study of thermodynamic properties of molecular fluids some methods such as integral equation method [10- 14], density functional theory [10, 15- 19] and computer simulation [10, 20- 24] have been used. Often, the molecular fluids have been studied in classical approximation [1]. In the present paper, we intend to extract a formalism based on the cluster expansion of the energy functional [25] for non-spherical molecular systems. Previously, this work has been done for systems with spherical particles which is called the lowest order constrained variational (LOCV) method [26]. Since this formalism does not depend on any free parameters in calculations, it is a self-consistent method. It considers the constraint in the form of a normalization condition which leads to the higher-order terms as small as possible. It also assumes a particular form for the long-range behavior of the correlation function in order to perform an exact functional minimization of the two-body energy with respect to the short-range behavior of the correlation functions [26]. Therefore, the LOCV formalism is a powerful technique for calculation of the thermodynamics of interacting many-particle systems, and it is preferable to other variational methods. The LOCV method has been extensively considered in the many-body calculation and investigation of interacting fermionic systems [27], many-nucleonic systems [28], normal and polarized liquid $^3$He

[29, 30]. In recent years, we have also applied this method to the study of various dense matter systems [28, 31] and the thermodynamic behavior of liquid $^3$He at zero and finite temperatures [32- 37]. As mentioned, in the present work, our purpose is to extract a formalism for non-spherical fluids. In other words, we intend to formulate the lowest order constrained variational calculation for studying the fluids with non-spherical molecules. We also examine the new formalism in the specific case of hydrogen fluid. The paper is arranged as follows. In section 2, we describe the LOCV formalism and extract a formulation for non-spherical molecules. We choose the Gay-Berne anisotropic potential for the interaction between molecules. We present this new formalism for hydrogen fluid in section 3. Conclusion is rendered in section 4.

## 2. FORMULATION OF THE LOWEST ORDER CONSTRAINED VARIATIONAL METHOD FOR NON-SPHERICAL MOLECULES

For an interacting system of particles, the best case of a variational method is to calculate the expectation value of the Hamiltonian using a trial wave function which includes the correlation between particles,

$$E = \frac{<\Psi|H|\Psi>}{<\Psi|\Psi>}, \qquad (8)$$

where $\Psi$ is the trial many-body wave function and defined as follows,

$$\Psi(1\dots N) = F(1\dots N)\Phi(1\dots N). \qquad (9)$$

$\Phi$ is the wave function of the system with no interaction, and $F$ is the correlation function of N-particles which actually induces the inter particles interaction effect in the wave function [25]. For $N$ particles in three dimensions, the exact solution of the above equation for the energy calculation requires solving an 3N-dimensional integral. For the large N, this is indeed impossible. In the case where the density of system is not very high and the correlations between the particles are short-range, the cluster expansion of energy is an efficient approximation [25]. The cluster expansion process begins with the definition of a generalized normalization integral,

$$I(\alpha) = <\Psi|exp[\alpha(H-T_F)]|\Psi>, \qquad (10)$$

where $T_F$ is the ground state energy of particles without interaction and $\alpha$ is a parameter. In this way, the energy is as follows,

$$E = T_F + \frac{\partial G(\alpha)}{\partial \alpha}|_{\alpha=0}, \qquad (11)$$

with the generation function $G(\alpha) = lnI(\alpha)$. After a series of complex mathematics, Eq. (11) can be rewritten as a sum of the single-particle and multi-particle energies,

$$E = E_1 + E_2 + E_3 + \dots \quad (12)$$

Due to the short-range behavior of the inter particle correlations, the three-body and higher cluster energy contributions are small [26, 33], and we can consider up to the two-body cluster term,

$$E = E_1 + E_2. \quad (13)$$

For a system with non-spherical molecules the one-body energy is given by

$$E_1 = \sum_i \varepsilon_i, \quad (14)$$

where $\varepsilon_i$ is the energy of single-particle state and can include the translational, rotational and vibrational energies, and it is obtained by solving the following eigenvalue equation,

$$t(i)\varphi_i = \varepsilon_i \varphi_i, \quad (15)$$

where $t(i)$ is the single-particle energy operator and $\varphi_i = \langle \vec{r}, \hat{u}|i \rangle$ is the single-particle wave function.

The contribution of two-body cluster energy is calculated as follows,

$$E_2 = \frac{1}{2}\sum_{ij} <ij|W(12)|ij>, \quad (16)$$

where $W(12)$ is the two-body effective potential which in fact in addition to the interaction potential, also has the particles correlation contribution,

$$W(12) = \frac{1}{2}F^\dagger(12)[t(1) + t(2), F(12)] + \frac{1}{2}[F^\dagger(12), t(1) + t(2)]F(12)$$
$$+F^\dagger(12)U(12)F(12). \quad (17)$$

$U(12)$ is the inter-particle potential that it is anisotropic for non-spherical molecules, which means that $U(12) = U(\vec{r}, \hat{u}_1, \hat{u}_2)$. $F(12)$ is the two-body correlation function. Here we consider the choice of correlation in the form of Jastrow ansatz, $F(1 \dots N) = \prod_{i<j} f(ij)$ [38]. Applying the Jastrow form, we have

$$F(12) = F^\dagger(12) = f(r). \quad (18)$$

With this choice, the rotation and vibration parts of $t(i)$ are commuted with $f(r)$ and Eq. (17) takes the following form,

$$W(12) = W(\vec{r}, \hat{u}_1, \hat{u}_2) = \frac{\hbar^2}{m}[f'(r)]^2 + f^2(r)U(\vec{r}, \hat{u}_1, \hat{u}_2), \tag{19}$$

where $m$ is the mass of the molecule and $\hbar = \frac{h}{2\pi}$ is the reduced Planck constant. As mentioned in the previous section, a non-spherical molecule has two characteristics in the space. Therefore, we can write the single-particle wave function of the system as,

$$\varphi(\vec{r}, \hat{u}) = \frac{1}{\sqrt{\Omega}} e^{i\vec{k}\cdot\vec{r}} Y_l^m(\hat{u}), \tag{20}$$

where $\vec{k}$ is the wave vector attributed to the molecule. Finally, by performing some algebra, we get the two-body energy equation as follows

$$E_2 = \frac{1}{2\Omega} \iiint d\vec{r} d\hat{u}_1 d\hat{u}_2 W(\vec{r}, \hat{u}_1, \hat{u}_2) a(\vec{r}, \hat{u}_1, \hat{u}_2), \tag{21}$$

where,

$$a(\vec{r}, \hat{u}_1, \hat{u}_2) = \sum_{\substack{k_i l_i \\ k_j l_j}} \frac{(2l_i+1)(2l_j+1)}{(4\pi)^2} \left[1 \pm e^{i(\vec{k}_i - \vec{k}_j)\cdot\vec{r}} P_{l_i}(\hat{u}_1 \cdot \hat{u}_2) P_{l_j}(\hat{u}_1 \cdot \hat{u}_2)\right], \tag{22}$$

where $\vec{k}_i$ and $\vec{k}_j$ are the wave vectors corresponding to the $i$-th and $j$-th states, respectively. The upper (lower) sign in Eq. (22) corresponds to the Bosonic system (Fermionic system) case.

To calculate the energy, we need to obtain the correlation function. In the LOCV formalism, this is made by considering the normalization constraint of the wave function and defining a parameter as smallness parameter [25]. For our system, this parameter will be as,

$$\xi = \frac{1}{N\Omega} \iiint d\vec{r} d\hat{u}_1 d\hat{u}_2 (f^2(r) - 1) a(\vec{r}, \hat{u}_1, \hat{u}_2). \tag{23}$$

Then, with functional minimization of two-body energy under normalization constraint and using the Euler-Lagrange equation [39], we get a differential equation as follows,

$$f''(r) + \left[\frac{2}{r} + \frac{A'(r)}{A(r)}\right] f'(r) - \frac{2m}{\hbar^2} \left[\frac{\iiint d\hat{u}_r d\hat{u}_1 d\hat{u}_2 U(\vec{r}, \hat{u}_1, \hat{u}_2) a(\vec{r}, \hat{u}_1, \hat{u}_2)}{A(r)} + \frac{\lambda}{N}\right] f(r) = 0, \tag{24}$$

where $\lambda$ is the Lagrange multiplier which imposes the normalization constraint, and

$$A(r) = \iiint d\hat{u}_r d\hat{u}_1 d\hat{u}_2 a(\vec{r}, \hat{u}_1, \hat{u}_2). \tag{25}$$

By solving the above differential equation, we obtain the correlation function.

One of the important parameters in the study of fluids is the distribution function. For a many-body system in cluster expansion, the two-body distribution function is obtained as follows [25],

$$g(12) = f^2(12) \sum_{n=2}^{N} [\Delta g(12)]_n. \tag{26}$$

For the system under consideration, in the first term, the two-body distribution function is given by

$$g(\vec{r}, \hat{u}_1, \hat{u}_2) = f^2(r) a(\vec{r}, \hat{u}_1, \hat{u}_2). \tag{27}$$

Here we see that by computing the correlation function, $f(r)$, the distribution function and later the thermodynamic properties of the system can be calculated.

As we know, studying the physical properties of a system depends highly on understanding the interactions between particles. Both in classical and quantum mechanics, it is very difficult to obtain the accurate solutions when we enter the inter particle interactions. So we should use the methods with the mathematical approximations. These are techniques based on mathematical estimations, for example, in single-particle approximations, the interaction effect is considered as a mean field. The Hartree-Fock method is based on this mean field method. In the perturbation theories, the interaction effect is introduced as a perturbation term. In the case of many-body systems with strong interactions, these methods are unable to lead the good results. Because a strong interaction cannot be introduced as a mean field or perturbation effect. In the study of molecular fluids, the study of system thermodynamics in the presence of interactions requires the solution of relatively complex equations that are usually simplified by hard core approximation in some methods. As it was mentioned in the previous sections, by applying these approximations, the results depart from observations. But the extracted formalism does not have this limitation, rather, it can be applied to any system with any type of interaction. This important preference, in addition to the advantages mentioned in the preceding sections for the LOCV formalism, its successful application in the study of spherical particles and the excellent compatibility of its results with experiment and simulation, lets us to formulate this approach for systems with non-spherical molecules.

For this matter, different anisotropic potentials [2, 20] can be selected for the interaction of two non-spherical molecules. We chose the anisotropic shape of the 12-6 Lennard-Jones potential, the Gay-Bern potential, with the advantages outlined in the previous section. This new formalism can be used to investigate fermionic or bosonic systems at zero and finite temperatures in bulk or low dimensional cases. In the present work, as an example, we apply the formulation obtained in this paper for the bulk hydrogen fluid at finite temperature in the next section.

## 3. APPLICATION OF THE FORMALISM FOR $H_2$ FLUID

$H_2$ fluid is formed by the hydrogen molecules which are diatomic and linear, therefore it is one of the non-spherical systems that can be studied using the LOCV formalism extracted in the previous section. In the present work, we study hydrogen fluid at finite temperature. We choose the temperature in the range below the excitation temperature of rotational and vibrational degrees of freedom of hydrogen molecules. Therefore, in the present calculations, the rotation and vibration of the molecules have no contribution to the energy. In other word, we consider the hydrogen molecules in the rotational and vibrational ground states, $l_i = m_i = l_j = m_j = 0$. Applying these conditions, we rewrite the formulation of previous section for a hydrogen fluid with the number density $\rho = \frac{N}{\Omega}$ at the temperature $T$. It should be noted that for the hydrogen molecules, the values of parameters taken in our calculations have been given in Table 1.

Table 1: The value of parameters for hydrogen molecules.

| $\frac{\hbar^2}{mk_B}$ $(KA^2)$ | $\sigma_0$ $(A)$ | $\frac{\epsilon_0}{k_B}$ $(K)$ |
|---|---|---|
| 24.06 | 2.93 | 37.0 |

For this system, one-body energy per particle is as follow,

$$\frac{E_1}{N} = \frac{\hbar^2}{4\pi^2 m\rho} \int_0^\infty n(k)k^4 dk, \qquad (28)$$

where $n(k)$ is Bose-Einstein distribution function,

$$n(k) = \frac{1}{e^{\beta(\varepsilon(k)-\mu)}-1}. \qquad (29)$$

In Eq. (29), $\mu$ is the chemical potential which is obtained by imposing the following constraint,

$$\rho = \frac{1}{2\pi^2} \int_0^\infty \frac{k^2 dk}{e^{\beta(\varepsilon(k)-\mu)}-1}. \qquad (30)$$

Using Eq. (21), we get the following equation for two-body energy per particle of H$_2$ fluid,

$$\frac{E_2}{N} = \frac{\rho}{32\pi^2} \iiint d\vec{r}d\hat{u}_1 d\hat{u}_2 [\frac{\hbar^2}{m} f'^2(r) + f^2(r)U(\vec{r},\hat{u}_1,\hat{u}_2)]a(r). \qquad (31)$$

For hydrogen fluid, from Eq. (23), we can write the following relation for the smallness parameter,

$$\xi = 4\pi\rho \int_0^\infty r^2 [f^2(r) - 1]a(r)dr, \qquad (32)$$

where

$$a(r) = 1 + \frac{1}{4\pi^4\rho^2} [\int_0^\infty n(k)j_0(kr)k^2 dk]^2. \qquad (33)$$

Here, $j_0$ is the zeroth-order Bessel function of the first kind [39]. By applying Eqs. (31), (32) and (24), we obtain a differential equation to compute the correlation function as follows,

$$f''(r) + \left[\frac{a'(r)}{a(r)} + \frac{2}{r}\right]f'(r) - \frac{m}{\hbar^2}\left[\frac{1}{(4\pi)^3}\iiint U(\vec{r},\hat{u}_1,\hat{u}_2)d\hat{r}d\hat{u}_1 d\hat{u}_2 + \lambda\right]f(r) = 0. \quad (34)$$

By solving Eq. (34), we get the two-body correlation function, $f(r)$. In Fig. 1, the correlation function has been shown for a specific value of temperature at three different densities. As this figure shows, with increasing density, the correlation function tends to unity faster. Therefore, the correlation length decreases with increasing density as shown in Fig. 2.

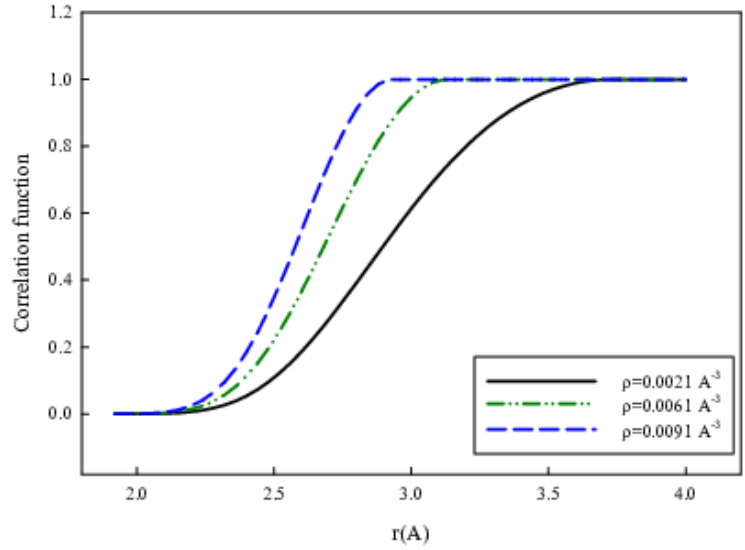

FIG. 1: Correlation function versus $r$ at $T = 3.5\ K$ and different densities ($\rho$) for fluid $H_2$.

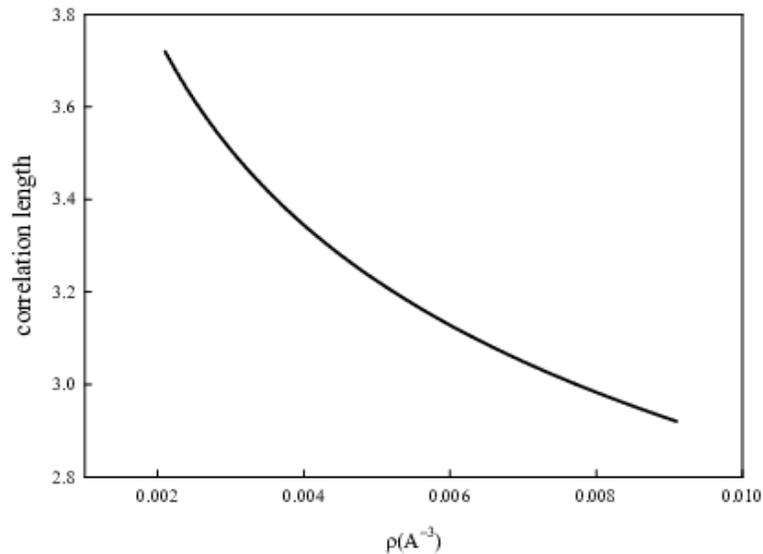

FIG. 2: Correlation length versus density at $T = 3.5\ K$ for fluid $H_2$.

The two-body distribution function for this system is,

$$g(r) = f^2(r)a(r). \tag{35}$$

In Fig. 3, we have presented the distribution function at a given temperature for different densities. Higher peaks in $g(r)$ are seen with increasing density at a fixed temperature. Also, as the distance between two molecules becomes large, $g(r)$ tends to unity. This means that at the higher intermolecular distances, the correlation between two molecules disappears.

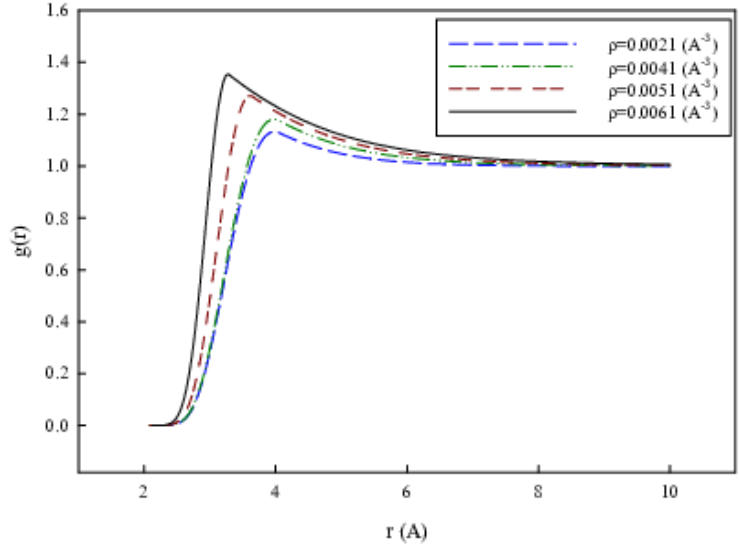

FIG. 3: Pair distribution function versus $r$ at $T = 3.5\ K$ and different densities ($\rho$) for fluid $H_2$.

The two-body static structure function, the counterpart of the momentum space of $g(r)$, is obtain by the following relation,

$$S(k) = 1 + 2\pi\rho \int_0^\infty \int_0^\pi [g(r) - 1] e^{i\vec{k}\cdot\vec{r}} \sin\theta d\theta r^2 dr. \tag{36}$$

This has been shown in Fig. 4 for some densities at a given temperature.

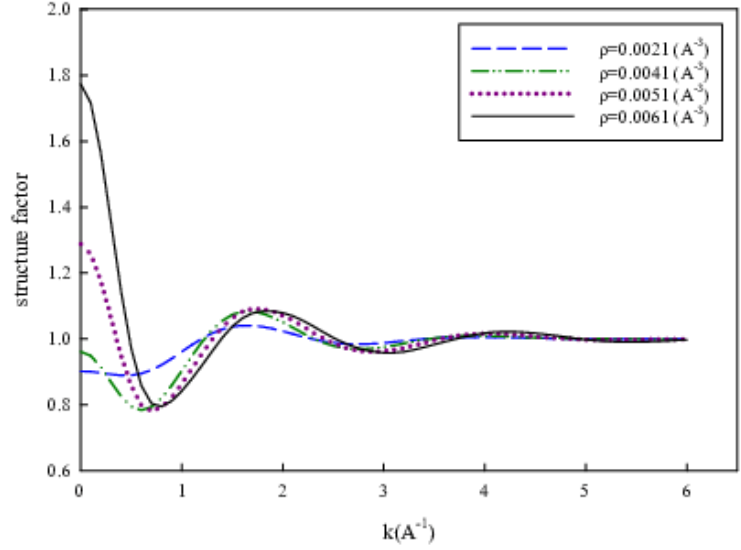

FIG. 4: Structure factor versus $k$ at $T = 3.5\ K$ and different densities ($\rho$) for fluid $H_2$.

Now, we calculate the heat capacity at constant volume from the following relation,

$$C_v = \left(\frac{\partial E}{\partial T}\right), \qquad (37)$$

where $E = E_1 + E_2$ is the internal energy of the system. In Fig. 5, we have drawn the heat capacity for the hydrogen fluid. As we expect for a bosonic system, the heat capacity increases with increasing temperature until it reaches its maximum value at condensation temperature, and thereafter, it decreases tending to its classical value. As the density increases, the peak of heat capacity becomes greater and shifts to the higher temperatures. This indicates that as the density increases, the Bose-Einstein temperature increases. In other word, the system is condensed at the higher temperatures.

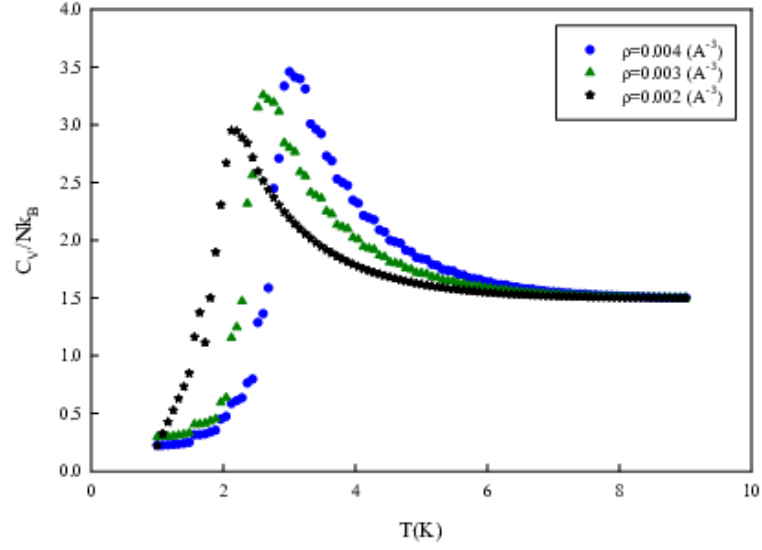

FIG. 5: Heat capacity versus temperature at different densities ($\rho$) for fluid $H_2$.

Entropy of the system is computed by applying the following relation [40],

$$\frac{S}{Nk_B} = \frac{1}{2\pi^2\rho} \int_0^\infty dk k^2 [(n(k)+1)\ln(n(k)+1) - n(k)\ln n(k)]. \qquad (38)$$

For hydrogen fluid, the entropy per particle has been plotted in Fig. 6 versus density. As it is expected, the entropy decreases by increasing the density, while it increases by rising the temperature. The temperature dependence of entropy before and after the Bose-Einstein density is different. This is well seen in Fig. 6.

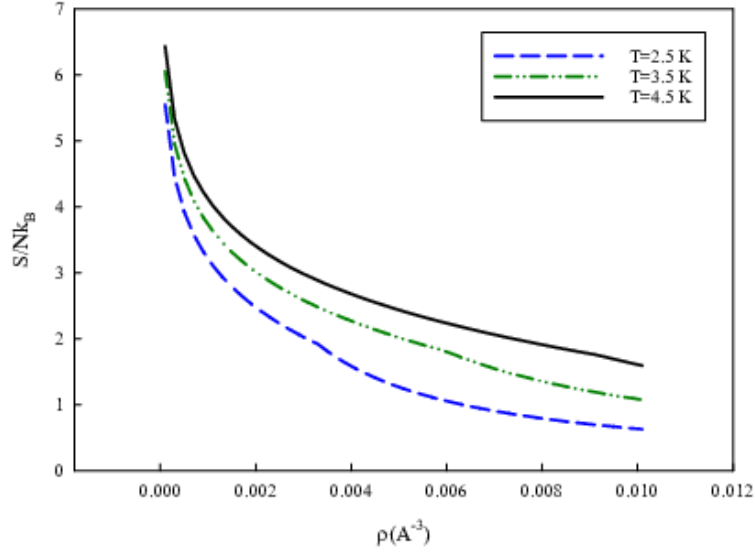

FIG. 6: Entropy versus density at different temperatures ($T$) for fluid $H_2$.

Helmholtz free energy is another important parameter in the thermodynamic investigation of a system. It can be obtained from the following relation,

$$F = E - TS. \tag{39}$$

Free energy per particle of hydrogen fluid versus temperature for three densities has been plotted in Fig. 7. This figure indicates that the free energy decreases with increasing the temperature. It also has larger negative values with increasing the density.

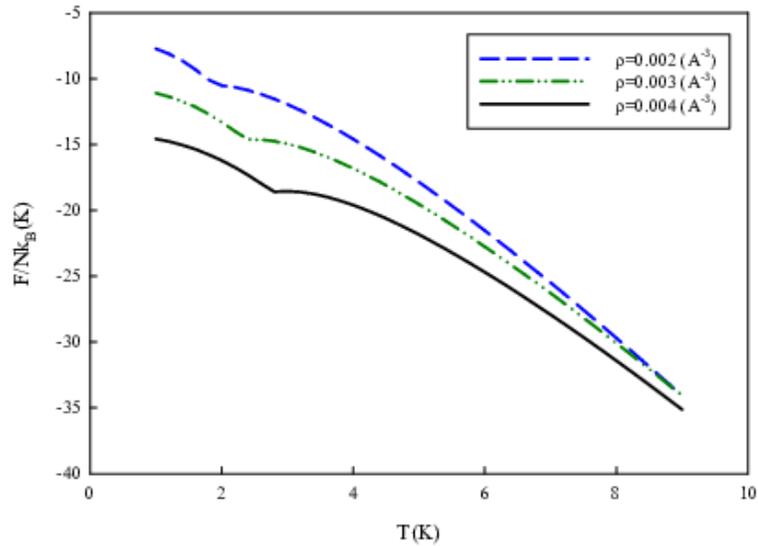

FIG. 7: Free energy per particle versus temperature at different densities ($\rho$) for fluid $H_2$.

Now, we can find the equation of state for the system using the following relation,

$$P = \rho^2 \frac{\partial F}{\partial \rho}. \tag{40}$$

In Fig. 8, the equation of state has been reported. As this figure shows, the pressure increases with increasing both the temperature and density.

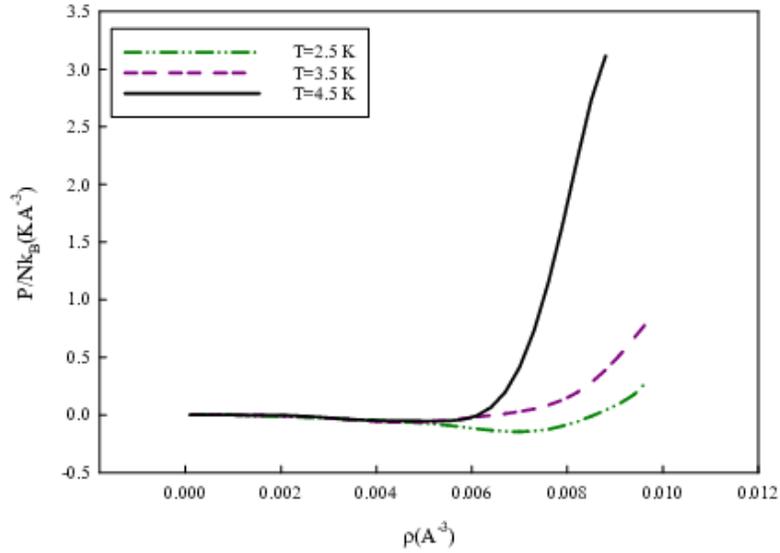

FIG. 8: Equation of state at different temperatures ($T$) for fluid $H_2$.

## 4. CONCLUSION

In summary, we have extracted a formalism for fluids with non-spherical molecules based on the cluster expansion of the energy functional. For this purpose, we have considered a system of linear molecules. For the interaction of two anisotropic molecules, we have applied the anisotropic form of the Lennard-Jones potential, the Gay-Berne potential. Also, we have exerted the Jastrow ansatz for the correlation function. Later, we have minimized the two-body energy term imposing the normalization constraint and obtained an Euler-Lagrange differential equation. By numerically solving this differential equation, the correlation function can be obtained for the system, then the thermodynamic properties of the system can be calculated. This new formalism has no restriction on the type of the interaction. In general, this formalism can be used for all systems, whether bosonic or fermionic at any temperature and with any type of interparticle potential. Here, as an example, we have applied this new formalism to calculate some thermodynamic properties of hydrogen fluid which confirmed some expected results.

Acknowledgments

We wish to thank Shiraz University Research Council. S. M. Hekamatzadeh wishes to thank M. A. Rastkhadiv for his useful comments and discussions during this work.